\documentclass[preprint,showpacs,preprintnumbers,amsmath,amssymb]{revtex4}
\usepackage{graphicx}% Include figure files
\usepackage{dcolumn}% Align table columns on decimal point
\usepackage{bm}% bold math

\begin{document}

%\preprint{APS/123-QED}

\title{Behavior of the contacts of Quantum Hall Effect devices \\ at high currents : an electronic thermometer}% Force line breaks with \\

\author{Y. M. Meziani,  C. Chaubet, S. Bonifacie, B. Jouault and A. Raymond}
\vspace{0.5cm}
\affiliation{ Groupe d'Etude des Semiconducteurs, UMR CNRS 5650, Universit\'e Montpellier II, 34095 Montpellier cedex France.} 
\author{W. Poirier and F. Piquemal}
\vspace{0.5cm}
\affiliation{ Bureau National de M\'etrologie, Laboratoire National d'Essais (BNM-LNE) 33, avenue du g\'en\'eral Leclerc 92260 Fontenay aux Roses, France} 

\date{\today}% It is always \today, today,
\begin{abstract}
\par This paper reports on an experimental study of the contact resistance of
Hall bars in the Quantum Hall Effect regime while increasing the current
through the sample. These measurements involve also the longitudinal
resistance and they have been always performed before the breakdown of the
Quantum Hall Effect. Our investigations are restricted to the $i=2$ plateau
which is used in all metrological measurements of the von Klitzing constant
$R_K$. A particular care has been taken concerning the configuration of the
measurement. Four configurations were used for each Hall bar by reversing the
current and the magnetic field polarities. Several samples with different
width have been studied and we observed that the critical current for the
contact resistance increases with the width of the Hall bar as previously
observed for the critical current of the longitudinal resistance. The critical
currents exhibit either a linear or a sublinear increase. All our observations are interpreted in the current understanding of the Quantum hall effect brekdown. Our analysis suggests that a heated region appears at the current contact, develops and then extends in the whole sample while increasing the current. Consequently, we propose to use the contact resistance as an electronic thermometer for the Hall fluid. 
\end{abstract}

\maketitle
%========================================
\section{Introduction}
%========================================
The Hall resistance $R_H$ of a two dimensional electron gas is quantized at
low temperature when the filling factor $\nu$ of the Landau levels is near an
integer~\cite{klit80,qhe87,tsui82}. Also, when the magnetic field is fixed at a
value corresponding to the center of the Hall plateaus, the longitudinal
resistance $R_{xx}$ of Hall-bar conductors vanishes as long as the current
does not exceed a critical value~\cite{klit80,qhe87,tsui82,robe89}. The
plateau $i=2$ is used in the metrological applications of the quantum Hall
effect (QHE) to provide a very reproducible resistance standard. Typically,
$R_H(i=2)$ does not deviate from $\frac{R_K}{2}$ by more than one part in
$10^9$ (relative value) if the Hall sample used fulfills certain
conditions. The von Klitzing constant $R_K$ is expected to correspond to the
ratio $\frac{h}{e^2}$. The recommanded value of  $R_K$ for metrology use is
$25812,807 \Omega$ with a relative uncertainty of one part in $10^7$~\cite{X}. The amplitude of the current which circulates across the sample must be limited to guarantee the accuracy of the measurement. Indeed, the onset of the longitudinal resistance $R_{xx}$ while increasing the current which is known as the breakdown of the Quantum Hall Effect~\cite{nachtwei,jeck,BNM}, destroys the total quantization of the system and prevents the measurement from being feasible. But in reality, far before the breakdown, another phenomenon affects the accuracy of the measurement of $R_K$. It is due to a linear relationship between $R_H$ and $R_{xx}$ ~\cite{Cage84,BNM}. Then, even a very small increase of the longitudinal resistance $R_{xx}$ can cause a deviation of $R_H$ from its expected value $\frac{R_K}{2}$. Typically, the deviation does not exceed one part in $10^9$ in relative value, only if the longitudinal resistance stays below $100 \mu \Omega$. This is the reason why the metrological measurements are performed using currents intensity which are always smaller than the breakdown current (see ~\cite{jeck} and ~\cite{BNM}). \\
\par In fact all metrological samples do not have the same limitation for the current. The four types of samples that have been studied for the aim of this work present marked differences. We have compared three PL sample series (PL173, PL174, PL175) to the LEP514 which is known for its optimal metrological qualities~\cite{dela}. All samples were grown and processed at the Philips Laboratory (PL formerly LEP) in Limeil Br\'evannes. The differences in the quality of the samples at high current denote a problem in the injection phenomenon of the electrons in the two dimensional electron gas. We then have to clarify the r\^ole of contacts in measuring the resistivity parameters in the QHE regime. We have also given great importance in our experiments, to the configuration of measure.\\

\par Of course the limitation of the current occurs before the breakdown of
the QHE and we will focus our attention on the current dependent properties in
the pre-breakdown regime, well before the abrupt onset of dissipation. Some
particularities of this regime are already known. First of all, it has been observed by many authors that the longitudinal resistance $R_{xx}$  exhibits an exponential increase as a function of the current ~\cite{eber83, komi85, bois94, jeck, fulr97}. \\
\par Another remarkable property of this regime has been reported in a recent publication by Kawano {\it et al.}~\cite{Kawano,Hisanaga} : it consists in an additional Cyclotron Emission (CE) signal observed in the vicinity of the source contact when the current is increased. The authors observed that at low-level current for which the two terminal resistance is still quantized, the cyclotron emission is observed at the electron entry and exit corners formed between the metallic current contacts and the two-dimensional electron gas. But when the current is increased, an additional CE signal is observed in the vicinity of the source contact. This is interpreted as the signature of a heating process that creates a non-equilibrium population of electrons near the contact. \\

\par Our paper reports on a third phenomenon : the abrupt increase of the
contact resistance at a current value which is lower or much lower than the
breakdown current. This phenomenon limits the current intensity that can be
injected in the Hall bars when doing metrological measurements. Our
experimental work is based only on transport measurements. We attribute the
increase of the contact resistance to the existence of a non equilibrium
electronic population created by a region of very high electric field near the
contact which injects the electrons in the sample. We discuss the correlation
of this abrupt increase of $V_c$ with the additional CE signal observed by Kawano et al in part IV.  \\

\par This paper is organized as follows. We present the GaAlAs/GaAs samples in part II and we detail our experimental protocol used to measure the sample voltage drop across the contact and the longitudinal voltage. Indeed it was important to identify the configuration that allows a general understanding of the injection mechanisms for any class of samples. The third part is devoted to the presentation of the experimental data which confirms that the configuration of the measurement is of prior importance. In the fourth part the results are discussed in the framework of the current understanding of the Quantum Hall Effect breakdown, and compared to other results.

%========================================
\section{Samples, experimental set up and protocol}
\subsection{Samples}
%========================================
\par The four types of samples investigated in this paper are GaAs/GaAlAs
heterojunctions grown on 3 inches wafers by Metal Organic Chemical Vapor
Deposition (MOCVD) technique. This technique allows to obtain a good
homogeneity of the electronic density over a large scale. Starting from the
substrate, a 600-nm-thick undoped GaAs buffer layer is firstly deposited. It
is followed by an undoped Al$_{0.28}$Ga$_{0.72}$As spacer layer whose thickness
is respectively 22 nm and 14.5 nm for PL175 and PL173 heterostructures. Then a 40-nm-thick $10^{18} cm^{-3}$ Si-doped Al$_{x}$Ga$_{1-x}$As layer is realized, with a gradual decrease of $x$ from 0.28 to 0 for PL175 and homogeneous $x$=0.28 value for PL173. The two other types of samples, PL174 and LEP514, have similar layers~\cite{poirierjap,BNM,piquemal}. Finally, an n-type 12 nm GaAs cap layer covers the heterostructure to improve the quality of ohmic contacts. After the realization of the 300-nm-thick delimiting mesa, the AuGeNi ohmic contacts are evaporated and then annealed at 450°C.\\
\par
\begin{figure}[htbp]
\begin{center}
\includegraphics[width=0.55 \textwidth]{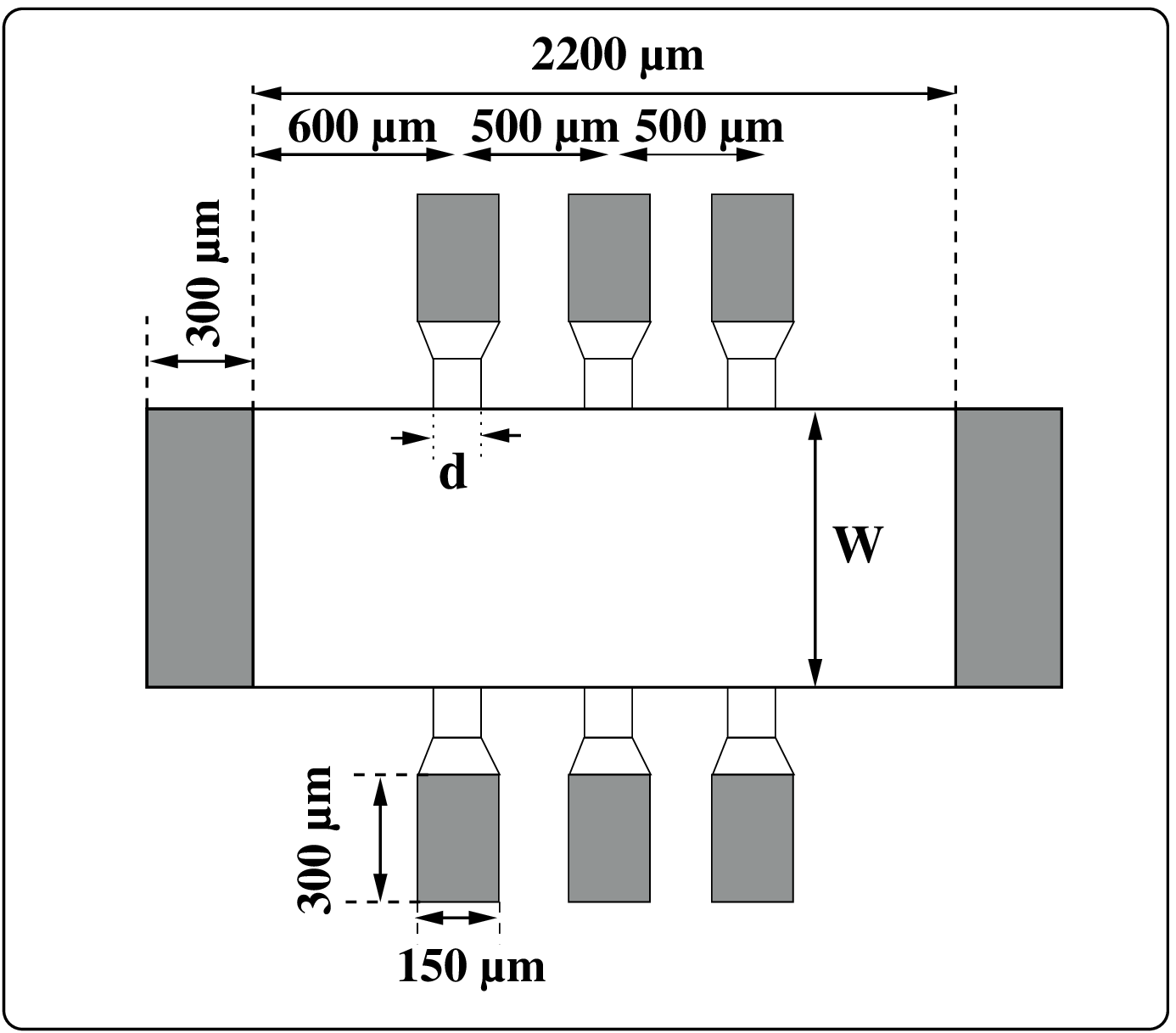}
\end{center}
\caption{The Hall bars are composed by a main channel whose width $W$ ranges
  from $200 \mu m$ to $1600 \mu m$. Two current contacts are used as a drain
  and a source. Six other lateral contacts which are smaller are used to
  measure the voltage drops. Their channel width d ranges from $5 \mu m$ to $150 \mu m$ as resumed in Table I.}
\label{fig:geo}
\end{figure}

 All samples were processed into a Hall bar. They have six
independent lateral contacts in addition to the source and drain contacts, as
described in figure~\ref{fig:geo}. LEP514 sample has only one pattern (see
table I). It was previously studied in an European Project~\cite{piquemal}. We
use it as a reference because of its recognized metrological qualities. The
other samples were patterned with different sizes. The Hall bar width $W$ of PL175
sample (respectively PL173) ranges from 200 to 800 (respectively 1600)
$\mu$m. These two series allowed us the study of the influence of the channel
width on the injection of electrons. The channel width $d$ of the lateral contact for samples PL174 ranges from 5 to 150 $\mu$m. This series has been studied using the lateral contacts located in the middle of the sample as a source and drain for electrons.\\
\begin{table}
\caption{\label{tab:table1} Samples characteristics ($\nu$ is the filling factor of the Landau levels)}
\begin{ruledtabular}
\begin{tabular}{cccccc}
WAFER & $N_s$ & $\mu$ & $B(\nu=2)$ &Widths $W$& length $d$\\
 & $(10^{15}m^{-2})$ &($\mathrm{m^2/V.s}$)&(T) & ($\mathrm{\mu m}$) & ($\mathrm{\mu m}$)\\
\hline
PL173& 3.3 & 50 & 6.8 T &200, 400 & 50\\
 & & & & and 1600 & \\
\hline
PL174& 4.5 & 50 & 9.4 T & 400 & 5, 50\\
 & & & & &and 150\\ 
\hline
PL175& 4.3 & 42.5 & 9 T &200, 400 & 50\\
 & & & & and 800 & \\
\hline
LEP514 & 5.1 & 30 & 10.7 T & 400 & 50\\
\end{tabular}
\end{ruledtabular}
\end{table}
\par The samples were connected on  TO-8 ceramic holders having 12 contacts and featuring leakage resistances between pins higher than 10$^{13} \Omega$ at room temperature. These ceramics have been mounted on a sample holder whose wires are $0.2  mm$ diameter constantan and placed inside a Variable Temperature Insert, in a 16 teslas superconducting magnet. All experiments were performed at 1.5K and 4.2K.

\subsection{Voltage drops near the contacts in the QHE regime}
%-------------------------------------------------------------
\par
\begin{figure}[htbp]
\begin{center}
\includegraphics*[width=0.55 \textwidth]{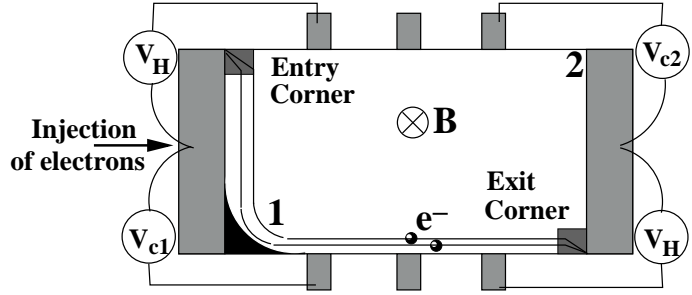}
\end{center}
\caption{Illustration of electron motion in the QHE regime and of the voltage drops in Hall bars. The resistance contact is measured in the opposite corners of the electron entry and exit corner referenced as points $1$ and $2$. The Hall voltage $V_H$ appears symetrically in the sample. The dark zone at point ``1'' represents the ``additional heated region''.}
\label{fig:configs}
\end{figure}

 In this part we detail the experimental protocol to measure the contact
resistance of the samples. All the measurements below have been carried out
only at the $i=2$ plateau. In the QHE regime, the magnetic field $B$ strongly
bends the potential profile. As a consequence, if a constant current $I$ is
applied between the source and drain contacts, all electrons enter the Hall
bar by one corner on the source side and leave by the opposite corner on the
drain side~\cite{klas, vanson, vanson2} (see figure~\ref{fig:configs}) . The existence of electron entry and
exit corners has been observed for the first time by Klass et al~\cite{klas}
and later by Kawano et al~\cite{Kawano}. These corners are high electric field
spots created by a concentration of the Hall potential over a very narrow
region of characteristic length $10\mu m$. This is illustrated in
figure~\ref{fig:configs}, where the entry and exit corners are clearly
labeled. The Hall voltage V$_H$ can be measured across those ``hot spots'' as shown
in figure~\ref{fig:configs}. One can measure the resistance of the current
contact through the other corners of the sample (corner labeled 1 and corner
labeled 2 on same figure) . \\
\par
\begin{figure}[htbp]
\begin{center}
\includegraphics*[width=0.55 \textwidth]{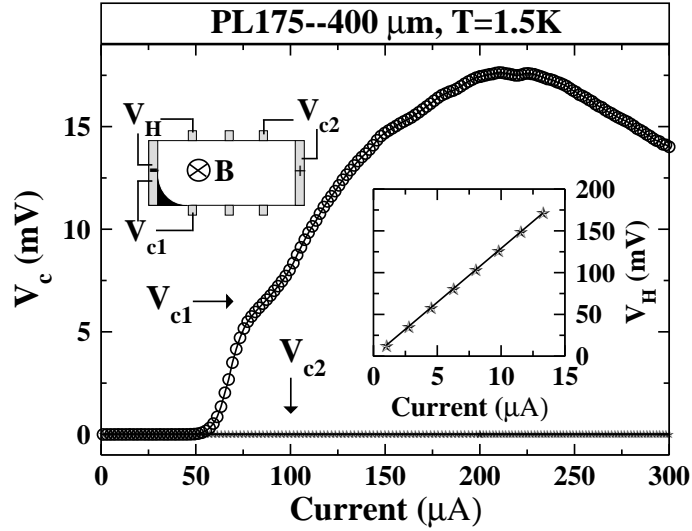}
\end{center}
\caption{Measurement of $V_{c1}$, $V_{c2}$ and $V_H$ for sample PL175 at the
  $i=2$ plateau. Electrons are injected from the left contact. In the inset, $V_H$ increases linearly with the current. $V_{c1}$ exhibits an abrupt onset of dissipation while $V_{c2}$ is stuck to zero. The heating process occurs only near the current contact which injects the electrons in the 2 DEG. The dark zone represents the ``additional heated region''. }
\label{fig:VccVh}
\end{figure}

 The voltage drop at the vicinity of the current contact is obtained using a
four-wires measurement. Two wires are used to inject the current in the Hall
bar and two other wires enable the measure of the voltage drop between one
current contact and one adjacent lateral contact. The Hall voltage $V_H= R_H
\times I$ may be measured between the two contacts surrounding either the
entry or the exit corner (see inset of figure~\ref{fig:VccVh}). More
surprising are the behaviors of both other corners labeled 1 and 2 which
differ one from another (Fig.~\ref{fig:configs}). This is obvious
in figure~\ref{fig:VccVh} where we reported the curves $V_{c1}(I)$ and
$V_{c2}(I)$ for sample PL175. The curve $V_{c1}(I)$
exhibits a steep increase at $I=50\mu$m, while $V_{c2}(I)$ remains stuck to
zero. $V_{c2}$ becomes finite only for higher current values corresponding
to the so-called breakdown regime. Since we are only interested in the onset
of dissipation, we will only focus on the contact whose resistance increases
the first. Then, $V_{c1}$ will be denoted $V_{c}$, $V_{c2}$ being
meaningless. A current polarity dependence in the measurement of longitudinal
voltages has already been reported by Komiyama {\it et al.}~\cite{komi96}. However
the geometry of their samples was completely different and particularly their
samples were narrower ($W=3 \mu m$). Hence, a direct comparison is not possible here. \\

\par  The abrupt increase of the contact resistance $V_c/I$ is the signature
of a heating process near the contact which injects the electrons in the
2DEG. We stress that the observation of Kawano {\it et al.}~\cite{Kawano} using Far
Infra Red (FIR) experiments, is another signature of the same heating phenomenon. As hot spots already exist in the sample, Kawano {\it et al.} called this new heating zone, the ``additional heated region''. We will come again to the discussion in part IV.

\subsection{Configurations of measure}
%-------------------------------------------------------------------------------------------
\par After passing through the ``heated region'', the hot electrons are
injected in the sample because of their drift velocity $V_d=\frac{E_H}{B}$
($E_H$ is the Hall electric field). As
explained in the bootstrap electron heating model
(BSEH)~\cite{komi96,komi85,kawa95,komi01,boel94,sagol}, the electrons propagate
their hot temperature. It is then appropriate to measure the longitudinal
voltage drops along the electrons paths. The distance between the current
contact and the first voltage probe is $600 \mu m$ (see
figure~\ref{fig:geo}) which is much larger than the minimum length which is
necessary to equilibrate the electronic temperature according to the BSEH
model. Komiyama et
al.~\cite{komi96} have measured a minimum length of $130 \mu m$ in the $i=2$ plateau to observe a steady state behavior of the longitudinal voltages. In our case, the geometry of our samples prevents us from observing such effects, so we will not be concerned by this length scale effect. \\
\par
\begin{figure}[htbp]
\begin{center}
\includegraphics*[width=0.55 \textwidth]{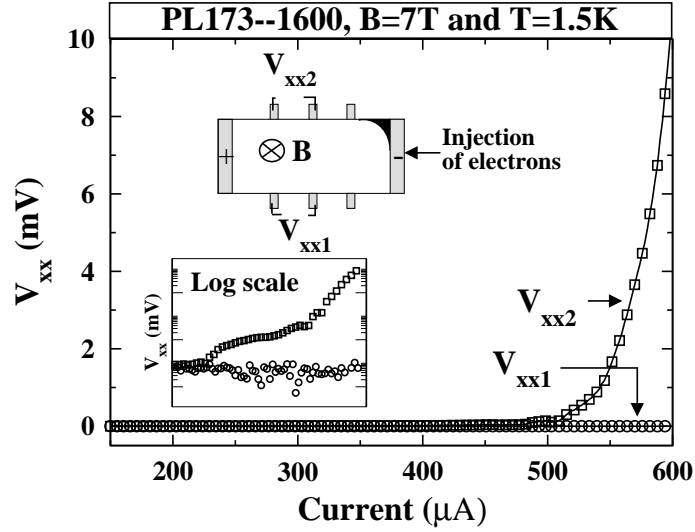}
\end{center}
\caption{Measurement of two longitudinal voltages $V_{xx1}$ and $V_{xx2}$ for
  sample PL173 with width Hall bar 1600 $\mu$m and for B=7T. $V_{xx2}$ is measured
  on the same side of the additional heated region along the electron flow. In the inset, the same
  graph in log scale highlights the huge difference in the behavior of each
  side of the sample (an offset has been added to $V_{xx1}$, the current scale is the same than in the figure).}
\label{fig:VxxVxx}
\end{figure}

 Following the works by Komiyama {\it et al.}~\cite{komi85} and by van Son et
al.~\cite{vanson}, the configuration for the measurement of the longitudinal resistance is of great importance. In our case, it was indeed of prior
necessity to measure the longitudinal voltage $V_{xx}$ which increases the
first. We have then compared the longitudinal voltage $V_{xx1}$ and $V_{xx2}$
measured simultaneously on lower and upper side of the sample (see
Fig.~\ref{fig:VxxVxx}). The voltage drop measured on the same side of the ``heated
region'' ($V_{xx2}$ in the inset) increases before the one
measured on the other side ($V_{xx1}$ in the inset). This observation has
been made on all the series of samples but was more obvious on the wide
samples. On the intermediate width samples ($W = 400 \mu m$) the discrepancy in the
behaviors of $V_{xx1}$ and $V_{xx2}$ was reduced but the longitudinal voltage
located in the continuity of the heated region always broke first while
increasing the current intensity. In the case of the narrower sample ($200 \mu
m$) the discrepancy was very small but still observable. This observation
highlights the r\^ole of the injection of electrons in the breakdown of the QHE
and justifies the care to be taken in the choice of the measurement configuration. \\

\par
\begin{figure}[htbp]
\begin{center}
\includegraphics*[width=0.55 \textwidth]{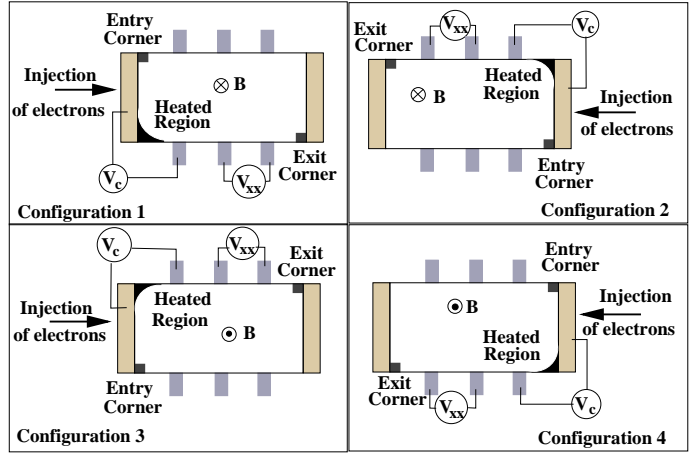}
\end{center}
\caption{Four configurations can be used to measure the contact voltage drop $V_c$ and the longitudinal voltage drop $V_{xx}$. These configurations are determined by the direction of the current and of the magnetic field.}
\label{fig:fourconfigs}
\end{figure}

 Reversing the current polarity and the magnetic field, one obtains four different configurations which are shown in figure~\ref{fig:fourconfigs}. Indeed if the current is reversed, then the electrons enter the opposite side of the Hall bar and the direction of the magnetic field determines where the hot spots appear. For all samples we studied the four configurations systematically.\\

%======================================================
\subsection{Experimental procedure}
%-----------------------------------------------------------------------------
\par
\begin{figure}[htbp]
\begin{center}
\includegraphics*[width=0.55 \textwidth]{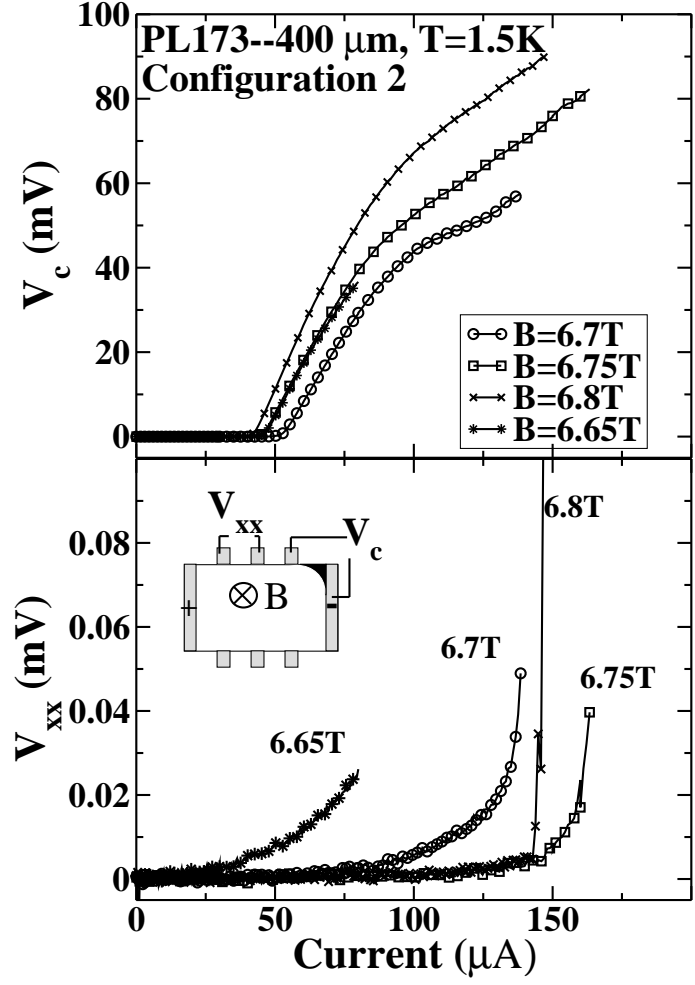}
\end{center}
\caption{Current dependence of $V_{xx}$ and $V_c$ for different magnetic fields around  $\nu=2$.}
\label{fig:VcVxx}
\end{figure}

 For each configuration 1, 2, 3 and 4 of figure~\ref{fig:fourconfigs}, our
experimental procedure is as follows. The $V_c(I)$ and $V_{xx}(I)$
characteristics are measured for a whole range of magnetic field around the
filling factor $\nu=2$. As an example, Fig.~\ref{fig:VcVxx} shows $V_c(I)$ and
$V_{xx}(I)$ at different filling factors. We define the current threshold of
one of these curves as the current for which the voltage reaches the value of
$50 \mu V$. It can be seen from figure~\ref{fig:VcVxx} that the current
threshold reaches a maximum value at a well defined value of the filling
factor $\nu$. Therefore, we can define $I_c$ and $I_b$ respectively as being
the maximum value for the $V_c(I)$ and $V_{xx}(I)$ current thresholds. \\

\par Finally, for a given configuration, one gets the value of $I_b$ and
$I_c$. While determining these currents, $V_{xx}$ was kept below the onset of
dissipation. Indeed, the reproducibility of the $V_c(I)$ and $V_{xx}(I)$ curves is not
satisfied anymore if the current is increased up to the
breakdown~\cite{boel94,ahle90}. Therefore we have focussed only on the
pre-breakdown regime in our study. Figure~\ref{fig:VcVxx} shows the
modifications induced by the magnetic field. We have noticed by sweeping the
magnetic field up and down, that the curves were reproducible. Several magnetic field values were investigated, allowing a precise and reproducible measurement of the critical currents $I_c$ and $I_b$ for each configuration.

%=================================================
\section{Experimental results}
%=================================================
\subsection{Variation of the critical currents with the channel width of the Hall bar.}
%-------------------------------------------------------
\par
\begin{figure}[htbp]
\begin{center}
\includegraphics*[width=0.55 \textwidth]{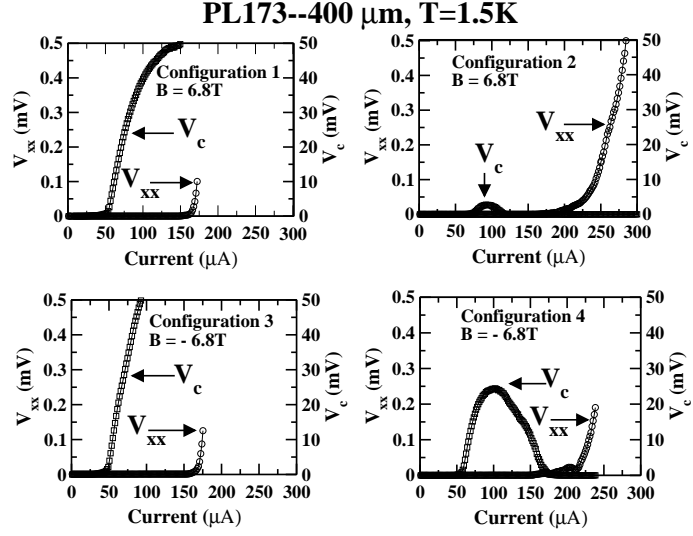}
\end{center}
\caption{Measurement of $V_{xx}(I)$ ('o')  and $V_c(I)$ ('$\square$') for
  sample PL173 with Hall bar width 400 $\mu m$ in the four
  configurations. These curves correspond to the magnetic field which gives the maximum values for $I_b$ and $I_c$. The critical current for $V_c$ is much lower than the critical current for $V_{xx}$. Configuration 1 (respectively 2) is similar of the configuration 3 (respectively 4).}
\label{fig:173400}
\end{figure}

 In this part, we present the results obtained for the PL173 and PL175
series. We have measured the critical currents for each sample in all
configurations following the experimental procedure described above. Before
discussing and comparing the results for both series of samples, we present as
an example the results obtained for the four configurations of PL173-400
sample. Figure~\ref{fig:173400} shows the  $V_c(I)$ and $V_{xx}(I)$
characteristics allowing the determination of $I_c$ and $I_b$ for this
sample. Each quadrant of Fig.~7 corresponds to one of the four configurations
given in Fig.~\ref{fig:fourconfigs}.\\

\par We must first notice that $I_c$ is systematically lower than $I_b$. The
second noticeable point is the strong similarity between the configurations 1
and 3 and the configurations 2 and 4. This is because in the configuration 1 and 3 electrons are injected by the
same current contact. In the configurations 2 and 4 electrons are injected by
the other one. We found that every contact of each sample
had its own signature. As a consequence, we took the mean value of the
critical currents obtained in configuration 1 and 3 (respectively 2 and 4) to
characterize one contact (respectively the other contact). We proceeded
differently for the threshold current $I_b$ because the measurement of $V_{xx}$ involves different pairs of lateral contacts in the sample. We charaterize $I_b$ with four different values corresponding to the four configurations. \\

\par
\begin{figure}[htbp]
\begin{center}
\includegraphics*[width=0.55 \textwidth]{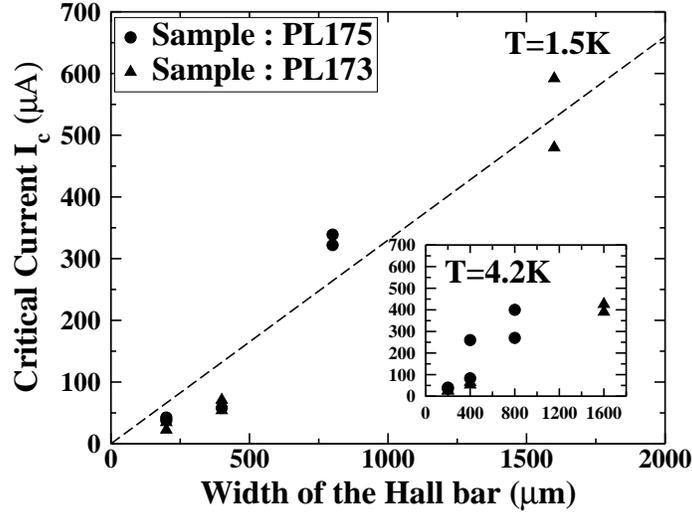}
\end{center}
\caption{The critical current $I_c$ of the contact resistance as a function of
  the width of the current contact (Hall bar). The dashed line is the linear fit $I_c=0.33 \times W$.}
\label{fig:IcW}
\end{figure}

 By these means, we could dress the graphs of the dependence of the
  critical currents with the channel width for the PL173 and PL175
  samples. Figure~\ref{fig:IcW} is devoted to the critical current $I_c$ as a
  function of the width of the Hall bar. We observe that the critical current
  increases with the width of the samples. This behavior is clearly observed
  for both PL173 and PL175 samples and besides we observe a
  linear increase. In such a way, the mean current density which is deduced
  from the data gives a mean value of $J_{cr}=0.33 A/m$ which can be compared
  to other results yet published. Measuring the breakdown of the QHE, many
  different values for $J_{cr}$ have been reported. We do not consider here
  the experiments which were done on laterally constricted samples because the
  values obtained in such cases are at least one order of magnitude above the
  usual ones. In the literature, for standard Hall bars, we found values
  ranging from $J_{cr}=0.5 A/m$ to $J_{cr}=1.6 A/m$~\cite{eber83,cage83,komi85,kawa94,nachtwei,jeck,okun95}.  \\

\par
\begin{figure}[htbp]
\begin{center}
\includegraphics*[width=0.55 \textwidth]{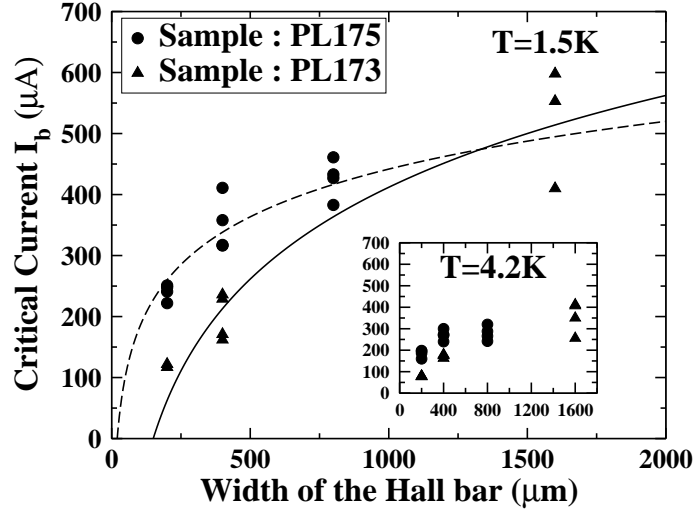}
\end{center}
\caption{The critical currents $I_b$ of $V_{xx}$ as function of the Hall bar
  width. Each point correspond to one configuration for each sample. Dashed
  line is the log fit : $I=500Log(W/150)$. Full line is the log fit : $I=260Log(W/25)$}
\label{fig:IbW}
\end{figure}

 We focus now on the general increase of $I_b$ with the width $W$. In
  figure~\ref{fig:IbW} the values of $I_b$ have been plotted as a function of
  the width $W$. There are four points for one sample. In this graph we
  clearly remark an important dispersion of the different points which
  characterize one sample. Again, this is an argument to stress that the
  configuration of measure is of prior importance when studying the heating
  effect in large Hall bars. The monotonic increase of $I_b$ is not obviously
  linear like the one of $I_c$. Indeed we could fit the results for $I_b$ with the law
  $I_{cr}=I_0Log(W/W_{0})$ stated by Balaban et al~\cite{bala93}. As shown in
  figure~\ref{fig:IbW}, the data for PL175 correspond to the curve
  $I=500Log(W/150)$ while the data for PL173 are fitted with
  $I=260Log(W/25)$. A sublinear increase for $I_b$ is not in contradiction with other works because the
  mobilities for the PL173 and 175 samples are quite high (see table I), and
  the possibility exists that the behavior should be not linear. In most
  samples studied in the literature the general tendancy for the
  breakdown current is a linear dependence $I_b(W)$ . Kawaji {\it et al.} \cite{kawa94, okun95} have already observed that the
  critical current of the breakdown $I_b$ increases linearly as a function of
  the Hall bar width. Also Boisen {\it et al.}~\cite{bois94} observed a linear
  increase of $I_b$. This law has only one exception which was reported by
  Balaban {\it et al.} who observed a sublinear increase of the critical current in
  very clean samples~\cite{bala93} whose mobility is $\mu =90$~$\mathrm{m^2/V.s}$. In another paper Balaban {\it et al.}~\cite{bala94} remarked that an intermediate mobility sample
  ($\mu = 12$~$\mathrm{m^2/V.s}$) could exhibit both a linear increase in the
  dark and a sublinear increase under illumination. Concerning our data, we can notice that the region of the contacts is undoubtly less perfect than the inner sample because of the inhomogeneities of the density caused by the presence of the ohmic contact. This can explain why $I_c(W)$ exhibits a linear increase, while $I_b(W)$ seems to fit with a sublinear increase.\\

\par Another clear tendancy is that the critical currents for PL173 samples are
  always lower than those for PL175 samples. We attribute this difference to
  the magnetic field at which the experiments are performed ($B = 9$~T
  for PL175, instead of $B=6.8$~T for PL173). Indeed, it was shown by Kawaji
  {\it et al.}~\cite{kawa94} and also by Jeckelmann {\it et al.}~\cite{jeck01} that the critical current of the
  longitudinal voltage scales with
  the magnetic field as $I_b \varpropto B^{\frac{3}{2}}$. This dependence is generally
  understood in terms of Inter Landau levels transitions~\cite{hein84,guim85,blie86,kirt86,blie88,eave86,kawa94}. \\

\par In both graphs on figure~\ref{fig:IcW} and ~\ref{fig:IbW}, the insets
  report the same measurements at the liquid helium temperature
  (T=4.2K). First these measurements confirm the measurements at 1.5K because
  the graphs are similar at first sight. Then, the modifications induced by a
  cooling of the system are clearly observable : most critical currents
  measured at 4.2K are lower than those measured at 1.5K. This increase of
  the critical current while lowering the temperature has been precisely studied by
  L.B. Rigal {\it et al.}~\cite{rigal}, although this was not for the critical current of the
  voltage drop across the contact.\\

%==============================================
\subsection{Comparison of the PL samples with the sample LEP 514}
%==============================================
\par In this part we compare the results obtained for the critical currents of
PL173, PL174 and PL175 samples to the critical currents of the sample
LEP514. In the previous part we have studied those critical currents as a
function of the width of the channel. But in order to compare the results with
those of the LEP514 whose channel width is $400$~$\mu$m, we will only consider the
PL173-400 and the PL175-400. Besides we can also make a comparison between the
LEP514 and the PL174(W=400 $\mu$m, d=50 $\mu$m) which have both exactly the same geometry (see table I). Using the current contact as a source, we
measured the values of the critical currents $I_c$ and $I_b$ for the LEP514
and the PL174-50. We reported in Table II, the mean value for the critical currents obtained for both contacts of each sample. \\
\begin{table}[htbp]
\caption{\label{tab:table2} Critical currents for the $400$~$\mu$m wide current contacts}
\begin{ruledtabular}
\begin{tabular}{ccccc}
SAMPLE& PL173 & PL175 & PL174 & LEP514\\
\hline
$I_c$& 75 & 70 & 120 & 440 \\
$I_b$& 200 & 350 &350  & 570  \\
\end{tabular}
\end{ruledtabular}
\end{table}
\par
\begin{figure}[htbp]
\begin{center}
\includegraphics*[width=0.55 \textwidth]{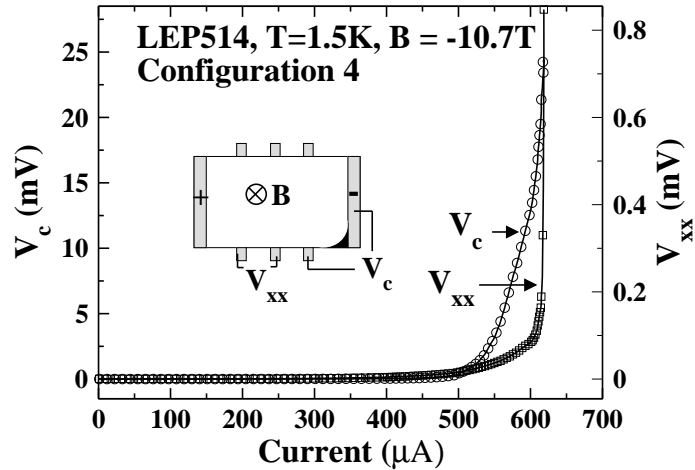}
\end{center}
\caption{Measure of $V_c(I)$ and $V_{xx}(I)$ for sample LEP514 at $i=2$ plateau for B=10.7T in configuration 4.}
\label{fig:lep514}
\end{figure}

 The differences between the LEP514 and the other three PL samples are
remarkable. First, the series PL exhibit always a much lower value for $I_c$
than for $I_b$. This is not the case for the LEP514. Second for LEP514, the
high values of these critical currents (around 500 $\mu$A) are much higher
than the values for the PL series (see Table II). LEP514 is known for allowing very accurate metrological measurements
of the Hall resistance in the Quantum Hall Effect regime~\cite{piquemal}. In
figure~\ref{fig:lep514}, we present the curves  $V_c(I)$ and $V_{xx}(I)$ for
one configuration of the LEP514 which allows one to observe that the critical
current of $V_c$ is equivalent to the critical current of $V_{xx}$. This
figure must be compared to Fig~\ref{fig:173400}. The particularities of this sample are certainly the cause of its reliability in allowing very accurate metrological measurements.\\

%=====================================
\subsection{Variation of the critical current with the channel width of the lateral contact.}
%=====================================
\par
\begin{figure}[htbp]
  \begin{center}
\includegraphics*[width=0.55 \textwidth]{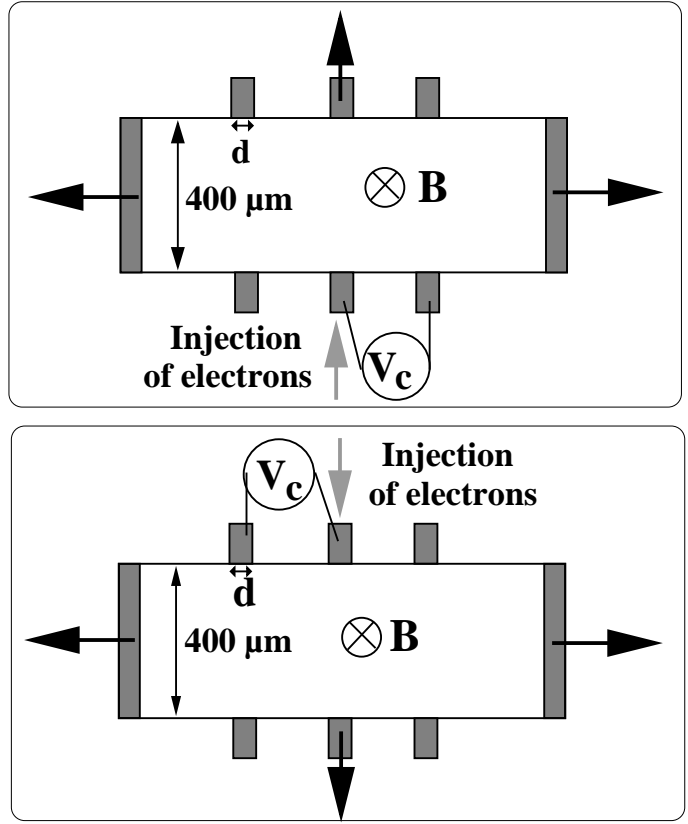}
\end{center}
\caption{Configurations of measurement of the contact resistance in the case of injection of electrons with the voltage probes}
\label{fig:configstension}
\end{figure}

\begin{figure}[htbp]
\begin{center}
\includegraphics*[width=0.55 \textwidth]{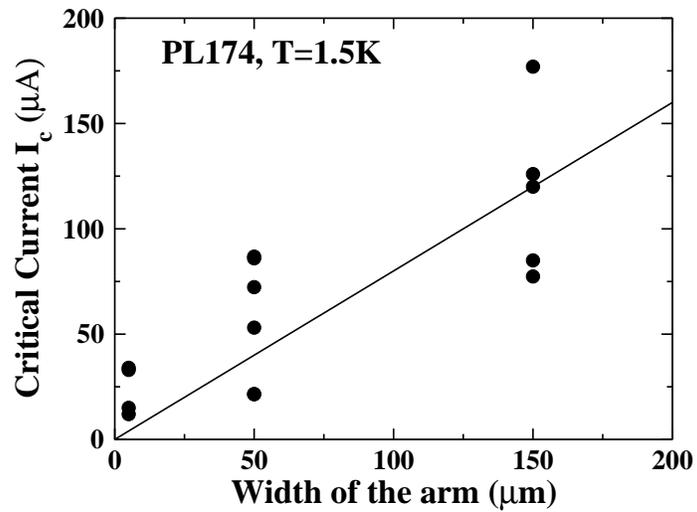}
\end{center}
\caption{Dependence of the critical current of V$_c$ for the voltage probes as a function of the width of the arm}
\label{fig:IcWtension}
\end{figure}

 In this part we discuss the influence of the width $d$ of the voltage probe arms
when using the lateral contacts as an electron source. For the PL174 series the
width of the Hall bar is fixed and equal to 400 $\mu$m. The channel width for
the voltage probes are 5 $\mu$m, 50 $\mu$m and 150 $\mu$m. We took as a source
the voltage contact located in the middle of the sample and the drain was
choosen in three possibilities : the two current contacts and the symetrical voltage
contact on the other side of the sample (see figure~\ref{fig:configstension}). For each
PL174 sample we studied both middle contacts by reversing the current. The
voltage drop $V_c$ was measured in the configurations shown in
figure~\ref{fig:configstension}. This method gave us six values for the
critical current $I_c$. These values are reported in
figure~\ref{fig:IcWtension}. The disparity observed in
figure~\ref{fig:IcWtension} in the values of the critical current for a given
width value is attributed to the contact itself and to the configurations which
seem to be not equivalent. This is again an argument to clearly define the
configuration of the measurements in such experiments. Nevertheless we observe an increase of the critical current
when increasing the width of the source contact, and despite the dispersion,
this behavior can be approximated by
a linear fit like for the critical current of the current contacts. The
resulting critical density is $J_{cr}=0.8 A/m$ exactly in the same range than
others results of the same type. More surprisingly, we observe that the
absolute values of the critical currents are of the same order of magnitude
than the ones for the current contacts. This is not expected because the width
of the channels for the voltage probes are much narrower ($150 \mu m$ at the
most). We cannot conclude anything here because of lack of data, but we can
nevertheless emphasize the problem of the influence of the voltage probes width in measuring the critical currents while using the current contacts as the source. Indeed, the similarity between the values obtained in completely different configurations is disturbing.  \\

%=====================================
\section{Discussion}
%=====================================

\par Throughout our experiments the PL samples exhibit a different behavior than the LEP514 : for PL samples of width $400 \mu m$ the critical currents $I_c$ were much lower than $I_b$, while in the case of the LEP514 the critical current $I_c$ and $I_b$ were nearly the same. We attribute the difference of the behaviors to the difference in the quality of the ohmic contacts or in the sample characteristics. However, although the behaviors of the samples have marked differences, we stress that the underlying physical mechanisms responsible for our observations are the same for all samples. Indeed, even in the case of the LEP514, $I_c$ is lower than $I_b$. In the following, we interpret the totality of our results in the framework of the current understanding of the Quantum Hall Effect breakdown. We show that there is evidence for the occurence of a large region of high electric field near the source, which can have different shape and dimension according to the sample characteristics. Nevertheless, we explain that the new experimental protocol we have proposed, allows us to highlight the particularities of any sample. For instance in the case the LEP514, which is actually known for its optimal metrological qualities, the heating region is small. Then we demonstrate that the resistance across the contact can be used as an electronic thermometer for the Hall fluid, if measured in the appropriate configurations presented in section II.C. This protocol migth be very useful for all breakdown experiments or measurements of resistance standards. \\

\subsection{High electric field domain near the source}
Our experimental findings demonstrate that the heating process appears first in the vicinity of the source contact before being visible in the middle of the sample. They demonstrate as well, that the mechanism of heating is different in both regions. We recall that $I_c$ scales linearly with the width $W$ of the sample, while $I_b$ scales sublinearly. Another difference between these two regions was highlighted in the experiments by Kawano {\it et al.}~\cite{Kawano} : the additional F.I.R. emission is visible only near the source, as previously described. All these remarks can be understood by the presence of an extended region of high electric field near the contact, as developed now.\\
Concerning first the onset of the contact resistance, it is obvious that this is the consequence of a QHE breakdown which occurs only near the contact. According to the present knowledge of this phenomenon, it is established that it has to develop in a sufficiently large domain of high electric field. In that case, the avalanche multiplication of electron hole pairs can occur along the drift of carriers, as well described by the BSEH model~\cite{komi01,sagol}. The abrupt jump of the voltage drop across the contact is then the consequence of the strong momentum exchanges between electronic states of the same Landau level. The interaction with the acoustical phonons is indeed enhanced by the presence of an intense electric field~\cite{chau98}. \\
Concerning now the Landau emission, which is the radiative part of the inter-Landau levels recombinations allowed by a non equilibrium Landau level occupation, it has been observed two different regimes of the FIR emission. A signal is first detectable at low current well before the breakdown. Klass {\it et al.}~\cite{klas} and Kawano {\it et al.}~\cite{Kawano} observed this far infrared signal emitted only by the entry and exit corners that they called hot spots. In fact it is now well understood that the non equilibrium population begins to appear when the drain source voltage exceeds half of the cyclotron energy ($eV_{SD}>\hbar\omega_c / 2$). In this regime, electrons can be injected directly from the reservoir to bulk electronic states in the upper Landau levels, leading to non-equilibrium population. Both tunneling between reservoirs and Landau levels and also between Landau levels, are efficient in this process. However in this regime, no threshold of $V_c$ can be observed because the high field zone is of restricted area and it does not affect the equilibrium of the edge states~\cite{Kawano}. Then, a second regime of emission appears at higher current values, but still before the breakdown in the sample. In the pre-breakdown regime, the FIR emission extends to the whole side of the source contact. The mechanism is different in that case because it leads to the increase of the contact resistance, but again the existence of an extended domain of high electric field can explain the onset of emission. Indeed the inter-Landau levels transitions assisted by acoustical phonons are enhanced by the strength of the electric field, and can cause the onset of a non equilibrium population of electrons, as described in details in ref.~\cite{chau95}. The occurence of both FIR emission and resistance onset in the vicinity of the contact, can be obtained for values of the electric field which are slightly over those required for the zener tunneling. \\

It is now clear that a high electric field domain of extended area is formed near the source, but we are still left with the question of the formation and the extension of this high field domain. Let us stress that the hot spot size increases drastically with the current amplitude in the pre-breakdown regime. First, the hot spot extends to the whole source side of the Hall bar. At even higher currents, it extends towards the drain side by following the electron flow. As a consequence, if the current is high enough, the contact used to measure $V_c$ is connected to a high electric field region. The extension has undoubtedly a different shape and dimension in different samples. For instance, in the case of the LEP514, the extension might be smaller than in the case of the PL samples, in which the high field domain reaches the edge of the sample where the measurement is done. The quality of the contacts certainly influences the formation of the high field domain, but the size of the sample is also an influent parameter. We recall that the currents $I_b$ and $I_c$ are approximately the same for the wide bars (see figures~\ref{fig:IcW} and ~\ref{fig:IbW}) and that the discrepancy observed in the measurement of $V_{xx1}$ and $V_{xx2}$ was much greater for wide bars. So in wide samples, the high field zone is certainly of restricted area compared to the total area of the sample. Contrarily, in narrow samples (when W is shorter than $200 \mu m$), the high field zone contaminates the Hall sample in a rather homogeneous way~\cite{shaskin} which explains why $V_{xx1}$ and $V_{xx2}$ are more similar.  These remarks allow us to understand why Klass {\it et al.}~\cite{klas} did not observe a ``heated zone'' at the opposite corner of the electron entry near the injection contact: their sample was very large ($W=2 mm$) and the high field zone is certainly less visible than in our narrow samples or in those of Kawano. \\ 

Before concluding this section we add two comments. Firstly, concerning the opposite side of the sample (the drain side). The reason why we do not observe an extended zone of high electric field is obviously due to the direction of propagation of electrons. On the drain side, if electrons are accelerated by the presence of a high field zone, they will enter the reservoir more rapidly, so reducing the possibility of the high field zone extension. Secondly, concerning what will happen when the breakdown in the entire 2DEG is reached. We propose in fact that this high electric field zone will extend afterwards to the entire edge of the sample, and might even leave the region of the contact. This would correspond to the observation made by Shaskin {\it et al.}~\cite{shaskin}, who clearly showed evidence that domains of high electric field move along the sample while the current amplitude is varied. However, even if there is evidence that these high electric fields regions play a key role in the onset of the breakdown~\cite{kaya}, the detailed physical mechanism of their extension and propagation remains to be developed.  

\subsection{The contact resistance as an electronic thermometer}
In consequence of the heating process near the source, the electronic temperature increases. Then, because of the drift velocity of the electrons, the electronic temperature is spread into the sample according to a mechanism which is well described by the BSEH model~\cite{komi96,kawa95}. Kaya {\it et al.}~\cite{kaya} showed evidence indeed of the slow relaxation of the hot electrons while they drift along Hall bars.\\

\begin{figure}[htbp]
\begin{center}
\includegraphics*[width=0.55 \textwidth]{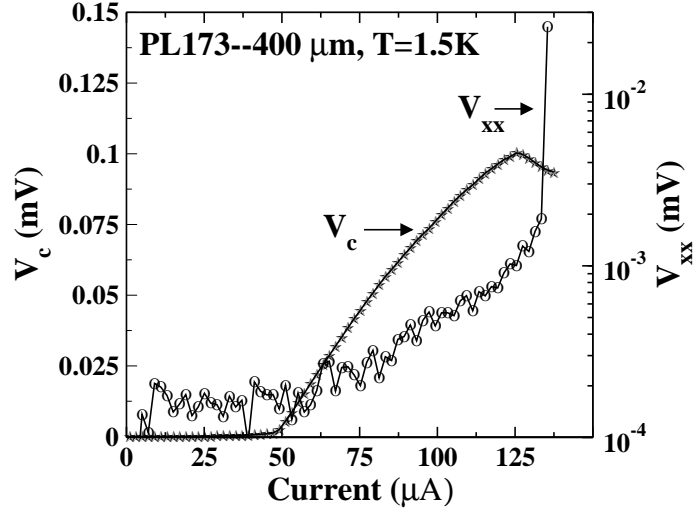}
\end{center}
\caption{Measurement of $V_{xx}$ (o) and $V_c$ ($\star$) for sample PL173 with
  Hall bar width 400 $\mu m$. $V_{xx}$ scale is logarithmic.}
\label{fig:Vxxlog}
\end{figure}

\begin{figure}[htbp]
\begin{center}
\includegraphics*[width=0.55 \textwidth]{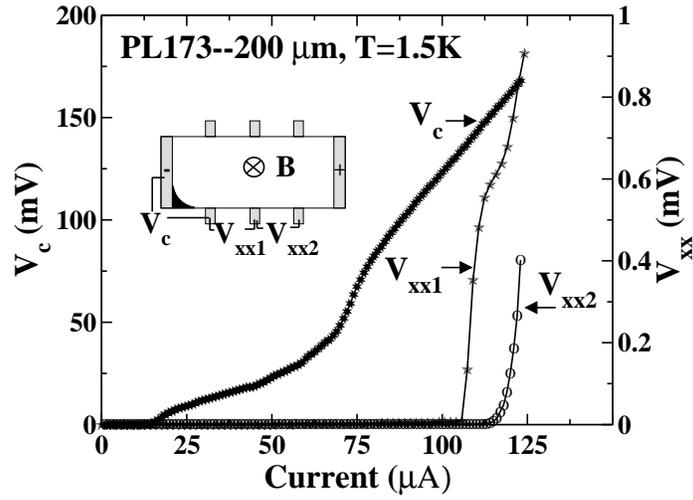}
\end{center}
\caption{$V_c$, $V_{xx1}$ and $V_{xx2}$ measured simultaneously for sample
  PL173 with Hall bar width of 200 $\mu$m in configuration 1.}
\label{fig:Vxx12Vc}
\end{figure}

To further pursue our investigation, we used our experimental results to highlight the correlation between the heating phenomenon near the contact and the transport properties of the sample far from the contact. In fact, we have measured simultaneously the voltage drop $V_c$ and the longitudinal voltage $V_{xx}$. We plotted the longitudinal voltage in a logarithmic scale to highlight the increase of this quantity which is very small in the pre-breakdown regime (see figure~\ref{fig:Vxxlog}). We clearly observe the exponential increase of the longitudinal conductivity with the current amplitude, which was reported several times by different authors~\cite{eber83,komi85,bois94} and refered to as the pre-breakdown regime. The point here is that it is apparent that this exponential increase of $V_{xx}$ begins with the onset of dissipation for $V_c$. This demonstrates that there exists a strong correlation between voltage drops at different places of the sample. It is also very interesting to notice that when $V_{xx}$ reaches the abrupt onset of dissipation (the breakdown), $V_c$ begins to decrease. The conclusion is that the heating process near the contact governs the transport mechanism along the electron flow. \\
Using specific Hall bars, Kaya {\it et al.} have shown that the hot electrons enter very far in the 2D system. Their experimental results are another manifestation of the same phenomenon, which is highlighted here. To prove this, we have also measured simultaneously (for one sample) the successive voltage drops along the Hall bar as in Kaya's experiments. These measurements are shown in figure~\ref{fig:Vxx12Vc}. As usual in narrow samples, the voltage drop across the contact increases first. But then, $V_{xx1}$ increases before $V_{xx2}$. Figure~\ref{fig:Vxx12Vc} demonstrates that the closer the investigated region is from the contact, the more influence the heating phenomenon has. Actually, Fig.\ref{fig:Vxx12Vc} demonstrates also that a second mechanism exists and is accountable for the increase of the longitudinal conductivity in the middle of the sample. If not, $V_{xx1}$ and $V_{xx2}$ would be the same. So, if it is clear that in the pre-breakdown regime the main effect is due to the increase of temperature near the source, it is also clear that a unique parameter is not enough to account for the increase of $V_{xx1}$ and $V_{xx2}$. Two explanations can be proposed to account for the observations. One explanation would be that the local electric field influences the transport phenomenon. A second explanation (proposed by Kaya {\it et al.}) would be to consider that a thermalization occurs while electrons are flowing away from the contact region, and that consequently the electron temperature reduces. \\
Nevertheless in both cases, there is evidence that the electronic temperature influences the transport through the entire structure, whatever the precise physical mechanisms involved here. As a consequence, we propose that the voltage drop $V_c$ can be used as an electronic thermometer for the Hall fluid in experiments on the breakdown of the QHE, and in metrological measurements as well. \\
Many additional experimental data and theoretical modeling would be necessary to calibrate this thermometer. We can nevertheless estimate roughly the value of this temperature in the different regimes. In the hot spots, well before the pre-breakdown regime, the electronic temperature has been estimated by Kawano {\it et al.} to be around $10$K. At higher current, in the heated region, the observation of the Landau emission is compatible with a temperature defined by~\cite{Kawano} $kT_e=\hbar \omega_c /7$ which leads in our case (for the PL173 at $\nu=2$, $\hbar \omega_c = 19 meV$ ) to $T_e=30$ K. In the case of Balaban's samples~\cite{bala93}, C. Chaubet {\it et al.}~\cite{chau98} have shown that this temperature increases exponentially from $10$K to $50$K at the threshold for longitudinal voltages. So it is clear that the electronic temperature is above $10$K in the pre-breakdown regime. However it is difficult to evaluate the maximum value which can be reached in the breakdown. By the way, let us remark that instabilities when reaching the breakdown can cause sudden high variation of the electric field and sudden increase of the electronic temperature. But a value of $30$K or even $50$K should not be very surprising at the onset of the breakdown because the Landau emission is observable and the energetic scales ($\hbar \omega_c$ and $kT_e$) should be comparable. However, let us remark that it is not highly probable that this temperature holds while electrons are propagating in the sample, because a thermalization (whose efficiency is not discussed here) occurs during the drift. The precise measurement of the electronic temperature would be of greatest interest at the onset of the breakdown, to determine which mechanisms are involved. In particular, it would be very interesting to know if the occurence of any universal value of $T_e$ can be demonstrated experimentally, or not. \\

%\newpage
\section{Conclusion}
\par A new measurement protocol has been used to characterize the pre-breakdown regime of the QHE. We have characterized for the first time the voltage drop across the contact and dressed the graph of the variation for its critical current with the width of the channel. After showing evidence that the usual longitudinal voltages has a different behavior, we have proposed that the contact resistance should be used as a thermometer for characterizing the temperature of the Hall fluid. The influence of $V_c$ on the pre-breakdown regime where the longitudinal conductivity increases exponentially with the current remains to be studied. Also, the propagation of the electronic temperature remains to be further studied in the context of the Bootstrap electron heating model. This work was granted by the Bureau National de M\'etrologie.\\

% This is how we recommend you do the references in LaTeX

%=====================Figures========================

\end{document}